\documentclass[12pt, draftclsnofoot, onecolumn]{IEEEtran}
\usepackage{amssymb}
\usepackage{mathtools}
\usepackage{graphicx}
\usepackage{soul}
\usepackage{cite}

\ifCLASSINFOpdf
\else
\fi
\begin{document}
%
\title{Extended Delivery Time Analysis for Non-work-preserving Packet Transmission in Cognitive Environment}
%
%
%

\author{Muneer Usman,~\IEEEmembership{Student Member,~IEEE,}
        Hong-Chuan Yang,~\IEEEmembership{Senior Member,~IEEE,}
        Mohamed-Slim Alouini,~\IEEEmembership{Fellow,~IEEE}
}

\maketitle

\begin{abstract}
Cognitive radio transceiver can opportunistically access the underutilized spectrum resource of primary systems for new wireless services. With interweave cognitive implementation, the secondary transmission may be interrupted by the primary user's transmission. To facilitate the packet delay analysis of such secondary transmission, we study the resulting extended delivery time that includes both transmission time and waiting time. In particular, we derive the exact distribution function of extended delivery time of a fixed-size secondary packet with non-work-preserving strategy i.e. interrupted packets will be retransmitted. Both continuous sensing and periodic sensing with and without missed detection cases are considered. Selected numerical and simulation results are presented for verifying the mathematical formulation. Finally, we apply the results to secondary queuing analysis with a generalized M/G/1 queue set-up. The analytical results will greatly facilitate the design of the secondary system for particular target application.
\end{abstract}


\section{Introduction}

\IEEEPARstart{R}{adio} spectrum resource scarcity is one of the most serious problems nowadays faced by the wireless communications industry. Cognitive radio is a promising solution to this emerging problem by exploiting temporal/spatial spectrum opportunities over the existing licensed frequency bands \cite{Goldsmith,Haykin,Mitola,Thomas,Akyildiz,Islam,Hamdaoui,Qianchuan,Qing}. Different techniques exist for opportunistic spectrum access (OSA). In underlay cognitive radio implementation, the primary and secondary users simultaneously access the same spectrum, with a constraint on the interference caused by the secondary user (SU) to primary transmission. With interweave cognitive implementation, the secondary transmission creates no interference to the primary user (PU). Specifically, the SU can access the channel only when the channel is not used by PU and must vacate the occupied channel when the PU appears. Spectrum handoff procedures are adapted for returning the channel to the PU and then re-accessing that channel or another channel later to continue/restart the secondary transmission. As such, the secondary transmission of a given amount of data may involve multiple transmission attempts and hence multiple spectrum handoffs, which results in extra transmission delay. The total time required for the SU to complete a given packet transmission will include the waiting periods before accessing the channel, the periods of the wasted transmissions, and the time used for the final successful transmission. In this paper, we investigate the statistical characteristics of the resulting extended delivery time (EDT) \cite{Borgonovo} and apply them to evaluate the delay performance of secondary transmission.

\subsection{Previous Work}

There has been a continuing interest in the delay and throughput analysis for secondary systems. For underlay implementation, \cite{Khan} analyzes the delay performance of a point-to-multipoint secondary network, which concurrently shares the spectrum with a point-to-multipoint primary network in the underlay fashion, under Nakagami-$m$ fading. The packet transmission time for secondary packets under PU interference is investigated in \cite{Sibomana}, where multiple secondary users are simultaneously using the channel. An optimum power and rate allocation scheme to maximize the effective capacity for spectrum sharing channels under average interference constraint is proposed in \cite{Musavian}. \cite{Tran} examines the probability density function (PDF) and cumulative distribution function (CDF) of secondary packet transmission time in underlay cognitive system. \cite{Farraj} investigates the M/G/1 queue performance of the secondary packets under the PU outage constraint. \cite{Jiang} analyzes the interference caused by multiple SUs in a ``mixed interweave/underlay'' implementation, where each SU starts its transmission only when the PU is off, and continues and completes its transmission even after the PU turns on.

For interweave implementation strategy, \cite{Gaaloul} discusses the average service time for the SU in a single transmission slot and the average waiting time, i.e. the time the SU has to wait for the channel to become available, assuming general primary traffic model. A probability distribution for the service time available to the SU during a fixed period of time was derived in \cite{Liang}. A model of priority virtual queue is proposed in \cite{J_Wang} to evaluate the delay performance for secondary users. \cite{Kandeepan} studies the probability of successful data transmission in a cooperative wireless communication scenario with hard delay constraints. A queuing analysis for secondary users dynamically accessing spectrum in cognitive radio systems was carried out in \cite{Li_Han}. \cite{Borgonovo} derives bounds on the throughput and delay performance of secondary users in cognitive scenario based on the concept of EDT. \cite{W_Wang} calculates the expected EDT of a packet for a cognitive radio network with multiple channels and users.

When the secondary transmission is interrupted by PU activities, the secondary system can adopt either non-work-preserving strategy, where interrupted packets transmission must be repeated \cite{Borgonovo}, or work-preserving strategy, where the secondary transmission can continue from the point where it was interrupted, without wasting the previous transmission \cite{W_Wang}. In our previous work \cite{Usman}, we carried out a thorough statistical analysis on the EDT of secondary packet transmission with work-preserving strategy, and then applied these results to the secondary queuing analysis. Typically, work-preserving packet transmission requires packets to be coded with certain rateless codes such as fountain codes, which may not be available in the secondary system.

\subsection{Contribution}

In this paper, we analyze the EDT or secondary packet transmission with non-work-preserving strategy, where the secondary transmitter needs to transmit the whole packet if the packet transmission was interrupted by PU activities. In general, the transmission of a secondary packet involves an interweaved sequence of wasted transmission slots and waiting time slots, both of which can have random time duration, followed by the final successful transmission slot. In this work, we first derive the exact expressions for the distribution function of EDT assuming a fixed packet transmission time. The generalization to random packet transmission time can be addressed in a similar manner as in \cite{Usman}. We consider three spectrum sensing scenarios -- i) ideal continuous sensing, in which the SU will continuously sense the channel for availability, ii) perfect periodic sensing, in which the SU will sense the channel periodically, and iii) imperfect periodic sensing, in which the SU will sense the channel periodically and there is a chance of sensing a free channel to be busy. For each scenario, we derive the exact statistics of the EDT for secondary packet transmission in terms of moment generating function (MGF) and PDF, which can be directly used to predict the delay performance of some low-traffic intensity secondary applications. To the best of our knowledge, the complete statistics of the EDT for non-work-conserving strategy has not been investigated in literature. We further apply these results to the secondary queuing analysis. Specifically, we investigate the queuing delay performance for the secondary system with periodic sensing in an M/G/1 setup. The queuing analysis for the other two sensing scenarios can be similarly addressed. The performance tradeoff involved in different sensing scenarios are investigated through selected numerical examples.

The rest of this paper is organized as follows. In section \ref{SystemModel}, we introduce the system model and the problem formulation. In section \ref{SecLT}, we analyze the EDT of a single secondary packet transmission for the three sensing scenarios. In section \ref{SecHT}, we analyze the average queuing delay of the secondary system in a general M/G/1 queuing set-up. This paper is finally concluded in section \ref{Conclusion}.

\section{System Model and Problem Formulation}
\label{SystemModel}

We consider a cognitive transmission scenario where the SU opportunistically accesses a channel of the primary system for data transmission. The occupancy of that channel by the PU evolves independently according to a homogeneous continuous-time Markov chain with an average busy period of ${\lambda}$ and an average idle period of ${\mu}$. Thus, the duration of busy and idle periods are exponentially distributed. The SU opportunistically accesses the channel in an interweave fashion. Specifically, the SU can use the channel only after PU stops transmission. As soon as the PU restarts transmission, the SU instantaneously stops its transmission, and thus no interference is caused to the PU.

\begin{figure} 
\includegraphics[width=6.4 in]{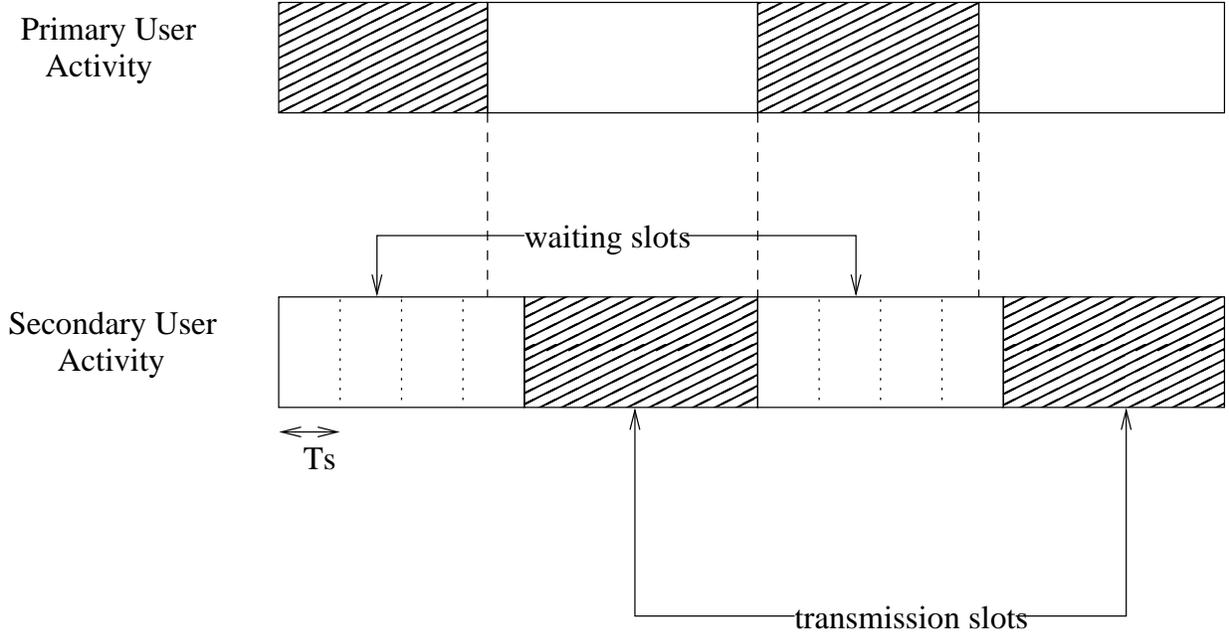}

\caption{Illustration of PU and SU activities and SU sensing for periodic sensing case.}
\label{cognitive_setup}
\end{figure}

The SU monitors PU activity through spectrum sensing. With continuous sensing, the SU continuously senses the channel for availability. Thus, the SU can start its transmission as soon as the channel becomes available. We also consider the case where the SU senses the channel periodically, with an interval of $T_s$. In particular, if the PU is sensed busy, the SU will wait for $T_s$ time period and re-sense the channel. With periodic sensing, there is a small amount of time when the PU has stopped its transmission, but the SU has not yet acquired the channel, as illustrated in Fig. \ref{cognitive_setup}. Under perfect periodic sensing scenario, the SU always senses correctly whether the channel is free or not. We also consider imperfect periodic sensing scenario, where there is a non-zero probability of missed detection, i.e. sensing a free channel to be busy in a certain sensing attempt. We assume that the chance of sensing a busy channel to be free is negligible, which can be achieved by adjusting the sensing thresholds properly. During transmission, the SU continuously monitors PU activity.  As soon as the PU restarts, the SU discontinues its transmission. The continuous period of time during which the PU is off and the SU is transmitting is referred to as a transmission slot. Similarly, the continuous period of time during which the PU is transmitting is referred to as a waiting slot. For periodic sensing case, the waiting slot also includes the time duration when the PU has stopped transmission, but the SU has not sensed the channel yet.

In this work, we analyze the packet delivery time of secondary system, which includes an interleaved sequence of the wasted transmission times and the waiting times, followed by a successful transmission time. Note that a transmission slot is wasted if its duration is less than the time required to transmit the packet. The resulting EDT for a packet is mathematically given by $T_{ED}=T_w+T_{tr}$, where $T_w$ is the total of the waiting time and wasted transmission times for the SU, and $T_{tr}$ is the packet transmission time. Note that both $T_w$ and $T_{tr}$ are, in general, random variables, with $T_w$ depending on $T_{tr}$, PU behaviour and sensing strategies, and $T_{tr}$ itself depending on packet size and secondary channel condition when available. Considering a fast varying channel and/or a long packet, the transmission time $T_{tr}$ can be estimated as a constant, given by \cite{Usman}
\begin{equation}
T_{tr} \approx \frac{H}{W \int_0^{\infty} \log_2(1+{\gamma})f_{\gamma}(\gamma) d{\gamma} },
\end{equation}
where $H$ is the entropy of the packet, $W$ is the available bandwidth, and $f_{\gamma}(\gamma)$ is the PDF of the SNR of the fading channel.
In what follows, we first derive the distribution of the EDT $T_{ED}$ for continuous sensing, perfect periodic sensing, and imperfect periodic sensing cases, which are then applied to the secondary queuing analysis in section \ref{SecHT}.

\section{Extended Delivery Time Analysis}
\label{SecLT}

In this section, we investigate the EDT of secondary system for a single packet arriving at a random point in time. These analyses also characterize the delay of some low-traffic-intensity secondary applications. For example, in wireless sensor networks for health care monitoring, forest fire detection, air pollution monitoring, disaster prevention, landslide detection etc., the transmitter needs to periodically transmit measurement data to the sink with a relatively long duty cycle. The EDT essentially characterizes the delay of measurement data collection.

\subsection{Continuous Sensing}
\label{SecLT_Cont}

The EDT for packet transmission by the SU consists of interweaved waiting slots and wasted transmission slots, followed by the final successful transmission slot of duration $T_{tr}$. We assume, without loss of generality, that the packet arrives at $t=0$. The distribution of $T_w$ depends on whether the PU was on or off at that instance. We denote the PDF of the waiting time of the SU for the case when PU is on at $t=0$, and for the case when PU is off at $t=0$, by $f_{{T_w},p_{on}}^{(c)}(t)$ and $f_{{T_w},p_{off}}^{(c)}(t)$, respectively. The PDF of the EDT $T_{ED}$ for the SU is then given by
\begin{equation}
f_{T_{ED}}^{(c)}(t) = \frac{\lambda}{{\lambda}+{\mu}} f_{{T_w},p_{on}}^{(c)}(t-T_{tr}) + \frac{\mu}{{\lambda}+{\mu}} f_{{T_w},p_{off}}^{(c)}(t-T_{tr}),
\label{ft_ed_con}
\end{equation}
where $\frac{\lambda}{{\lambda}+{\mu}}$ and $\frac{\mu}{{\lambda}+{\mu}}$ are the stationery probabilities that the PU is on or off at $t=0$, respectively. The two probability density functions $f_{{T_w},p_{on}}(t)$ and $f_{{T_w},p_{off}}(t)$ above are calculated independently as follows.

let ${{\cal{P}}_k}$ be the probability that the SU was successful in sending the packet in the $k^{th}$ transmission slot. This means that each of the first $(k-1)$ slots had a time duration of less than $T_{tr}$, while the $k^{th}$ transmission slot had a duration more than $T_{tr}$. Thus, ${{\cal{P}}_k}$ can be calculated, while noting that the duration of secondary transmission slots is exponentially distributed with mean $\mu$, as 
\begin{equation}
{{\cal{P}}_k} = e^{-\frac{T_{tr}}{\mu}} \cdot {\left(1 - e^{-\frac{T_{tr}}{\mu}}\right)}^{k-1}.
\label{P_k}
\end{equation}
For the case when PU is off at $t=0$, if a certain packet is transmitted completely in the $k^{th}$ transmission slot, then the total wait time for that packet includes $(k-1)$ secondary waiting slots and $(k-1)$ wasted transmission slots. Note that the duration of each of these $(k-1)$ waiting slots, denoted by the random variable $T_{wait}$, which is equal to PU on time, follows an exponential distribution for the continuous sensing case, with PDF given by
\begin{equation}
f_{T_{wait}}^{(c)}(t) = \frac{1}{\lambda}e^{\frac{-t}{\lambda}} u(t),
\label{ft_lam}
\end{equation}
while the duration of each of the previous $(k-1)$ wasted secondary transmission slots, denoted by the random variable $T_{waste}$, follows a truncated exponential distribution, with PDF given by
\begin{equation}
f_{T_{waste}}(t) = \frac{1}{1 - e^{-\frac{T_{tr}}{\mu}}} {\frac{1}{\mu}e^{\frac{-t}{\mu}}} \cdot (u(t) - u(t-T_{tr})),
\label{ft_mu_tr}
\end{equation}
where $u(t)$ is the unit step function. The MGF of $T_{w,p_{off}}$ for the continuous sensing case, ${\cal{M}}_{{T_w},p_{off}}^{(c)}(s)$ can be calculated as
\begin{align}
{\cal{M}}_{{T_w},p_{off}}^{(c)}(s) = \sum_{k=1}^{\infty} {{\cal{P}}_k} \times \left({\cal{M}}_{T_{wait}}^{(c)}(s) \right)^{k-1} \times \left({\cal{M}}_{T_{waste}}(s) \right)^{k-1},
\label{ftw_poff_con_summation}
\end{align}
where ${\cal{M}}_{T_{wait}}^{(c)}(s)$ is the MGF of $T_{wait}$ for the continuous sensing case, given by
\begin{equation}
{\cal{M}}_{T_{wait}}^{(c)}(s) = \frac{1}{1 - \lambda s },
\label{M_ft_cont_lam_k}
\end{equation}
and ${\cal{M}}_{T_{waste}}(s)$ is the MGF of $T_{waste}$, given by
\begin{equation}
{\cal{M}}_{T_{waste}}(s) = \frac{1 - e^{T_{tr}(s-\frac{1}{\mu})} }{(1 - \mu s )(1-e^{-\frac{T_{tr}}{\mu}})}.
\label{M_ft_mu_tr}
\end{equation}
After substituting Eqs. (\ref{P_k}), (\ref{M_ft_cont_lam_k}), and (\ref{M_ft_mu_tr}) into Eq. (\ref{ftw_poff_con_summation}), and applying the definition of binomial expansion on $({e^{T_{tr}(s-\frac{1}{\mu})} -1})^{k-1}$, Eq. (\ref{ftw_poff_con_summation}) becomes
\begin{equation}
{\cal{M}}_{{T_w},p_{off}}^{(c)}(s) = e^{-\frac{T_{tr}}{\mu}} \sum_{k=1}^{\infty} \frac{1}{(\lambda s - 1)^{k-1}(\mu s - 1)^{k-1}} \sum_{i=0}^{k-1} (-1)^i {{k-1}\choose{i}} \cdot {e^{i T_{tr}(s-\frac{1}{\mu})}}.
\end{equation}
Changing the sequence of the two summations, and applying the definition of negative binomial distribution, we get
\begin{equation}
{\cal{M}}_{{T_w},p_{off}}^{(c)}(s) = e^{-\frac{T_{tr}}{\mu}} + e^{-\frac{T_{tr}}{\mu}} \left[1 - {e^{T_{tr}(s-\frac{1}{\mu})}} \right] \sum_{i=0}^{\infty} (-1)^i {e^{i T_{tr}(s-\frac{1}{\mu})}} {\frac{1}{[s (\lambda \mu s - \lambda - \mu)]^{i+1}}}.
\end{equation}
Using the following general formula for partial fractions
\begin{equation}
\frac{1}{[x(x-a)]^n} = \sum_{j=0}^{n-1} (-1)^{n} {{2n-j-2} \choose {n-1}} \frac{1}{a^{2n-j-1}} \left[ \frac{1}{x^{j+1}} + \frac{(-1)^{j+1}}{(x-a)^{j+1}} \right],
\label{form_partial_fractions}
\end{equation}
the proof of which is given in the appendix, we get
\begin{equation}
{\cal{M}}_{{T_w},p_{off}}^{(c)}(s) = e^{-\frac{T_{tr}}{\mu}} - e^{-\frac{T_{tr}}{\mu}} \left[1 - {e^{T_{tr}(s-\frac{1}{\mu})}} \right] \sum_{i=0}^{\infty} \frac{{e^{i T_{tr}(s-\frac{1}{\mu})}}}{(\lambda \mu)^{i+1}} \sum_{j=0}^{i} {{2i-j} \choose {i}} \frac{1}{\alpha^{2i-j+1}} \left[ \frac{1}{s^{j+1}} + \frac{(-1)^{j+1}}{(s-\alpha)^{j+1}} \right],
\label{eq_A}
\end{equation}
where $\alpha=\frac{1}{\lambda} + \frac{1}{\mu}$. Taking the inverse MGF, and applying the definition of generalized hypergeometric function, we obtain the PDF of $T_{w,p_{off}}$ for continuous sensing case, as
\begin{align}
\nonumber
f_{{T_w},p_{off}}^{(c)}(t) = e^{-\frac{T_{tr}}{\mu}} \delta(t) + & \frac{e^{-\frac{T_{tr}}{\mu}} }{\lambda + \mu} (1 - e^{-\alpha t}) u(t) - \frac{e^{-\frac{2 T_{tr}}{\mu}} }{\lambda + \mu} (1 - e^{-\alpha(t-T_{tr})}) u(t-T_{tr}) \\
\nonumber
+ \sum_{i=1}^{\infty} \frac{(\lambda \mu)^i}{(\lambda + \mu)^{2i+1}}  {{2i} \choose {i}}  & \left[ {{}_1 F_1 \left(-i;-2i;-\alpha(t-i T_{tr}) \right)} e^{-(i+1) \frac{T_{tr}}{\mu}} u(t-i T_{tr}) \right. \\
\nonumber
& \left. - {{}_1 F_1 \left(-i;-2i;-\alpha(t-(i+1)T_{tr}) \right)} e^{-(i+2) \frac{T_{tr}}{\mu}} u(t-(i+1)T_{tr}) \right. \\
\nonumber
& \left. - {{}_1 F_1 \left(-i;-2i;\alpha(t-i T_{tr}) \right)} e^{-\alpha t} e^{-(i+1) \frac{T_{tr}}{\mu}} u(t-i T_{tr}) \right. \\
& \left. + {{}_1 F_1 \left(-i;-2i;\alpha(t-(i+1)T_{tr}) \right)} e^{-\alpha t}e^{-(i+2) \frac{T_{tr}}{\mu}} u(t-(i+1)T_{tr}) \right],
\label{ftw_cont_poff_final}
\end{align}
where ${}_1 F_1(.,.,.)$ is the generalized Hyper-geometric function. Note that the impulse corresponds to the case that the packet is transmitted without waiting.

For the case when PU is on at $t=0$, the MGF of $T_{w,p_{on}}$ for the continuous sensing case ${\cal{M}}_{{T_w},p_{on}}^{(c)}(s)$ can be similarly calculated as
\begin{align}
{\cal{M}}_{{T_w},p_{on}}^{(c)}(s) = \sum_{k=1}^{\infty} {{\cal{P}}_k} \times \left({\cal{M}}_{T_{wait}}^{(c)}(s) \right)^{k} \times \left({\cal{M}}_{T_{waste}}(s) \right)^{k-1}.
\label{ftw_pon_con_summation}
\end{align}
Using similar manipulations used for $T_{w,p_{off}}$, it is easy to arrive at
\begin{equation}
{\cal{M}}_{{T_w},p_{on}}^{(c)}(s) = -e^{-\frac{T_{tr}}{\mu}} \sum_{i=0}^{\infty} (-1)^i {e^{i T_{tr}(s-\frac{1}{\mu})}} \left[ {\frac{\mu s - 1}{[s (\lambda \mu s - \lambda - \mu)]^{i+1}}} \right].
\label{eq_B}
\end{equation}
Substituting Eq. (\ref{form_partial_fractions}) into Eq. (\ref{eq_B}), and carrying out some manipulation, we get
\begin{equation}
{\cal{M}}_{{T_w},p_{on}}^{(c)}(s) = e^{-\frac{T_{tr}}{\mu}} \sum_{i=0}^{\infty} \frac{{e^{i T_{tr}(s-\frac{1}{\mu})}}}{(\lambda \mu)^{i+1}} \sum_{j=0}^{i} {{2i-j} \choose {i}} \frac{\mu s - 1}{\alpha^{2i-j+1}} \left[ \frac{1}{s^{j+1}} + \frac{(-1)^{j+1}}{(s-\alpha)^{j+1}} \right].
\end{equation}
Performing some further manipulations, taking inverse MGF, and applying the definition of generalized hypergeometric function, we obtain the PDF of $T_{w,p_{on}}$ as
\begin{align}
\nonumber
& f_{{T_w},p_{on}}^{(c)}(t) = \frac{e^{-\frac{T_{tr}}{\mu}} }{\lambda + \mu} (1 + \frac{\mu}{\lambda} e^{-\alpha t}) u(t)\\
\nonumber
& + \sum_{i=1}^{\infty} \frac{(\lambda \mu)^i}{(\lambda + \mu)^{2i+1}}  \left[   {{2i} \choose {i}} {{}_1 F_1 \left(-i;-2i;-\alpha(t-i T_{tr}) \right)} \cdot e^{-(i+1) \frac{T_{tr}}{\mu}} \cdot u(t-i T_{tr}) \right. \\
\nonumber
& \left. -   {{2i} \choose {i}} \frac{\mu}{\lambda}  e^{-\alpha(t-i T_{tr})} {{}_1 F_1 \left(-i;-2i;\alpha(t-i T_{tr}) \right)} \cdot e^{-(i+1) \frac{T_{tr}}{\mu}} \cdot u(t-i T_{tr}) \right. \\
\nonumber
& \left. -   {{2i-1} \choose {i}} \left(1+\frac{\mu}{\lambda} \right) {{}_1 F_1 \left(1-i;1-2i;-\alpha(t-i T_{tr}) \right)} \cdot e^{-(i+1) \frac{T_{tr}}{\mu}} \cdot u(t-i T_{tr}) \right. \\
& \left. +   {{2i-1} \choose {i}} \left(1+\frac{\mu}{\lambda} \right) e^{-\alpha(t-i T_{tr})} {{}_1 F_1 \left(1-i;1-2i;\alpha(t-i T_{tr}) \right)} \cdot e^{-(i+1) \frac{T_{tr}}{\mu}} \cdot u(t-i T_{tr}) \right].
\label{ftw_cont_pon_final}
\end{align}

\begin{figure}[htb]
\includegraphics[width=6.4 in] {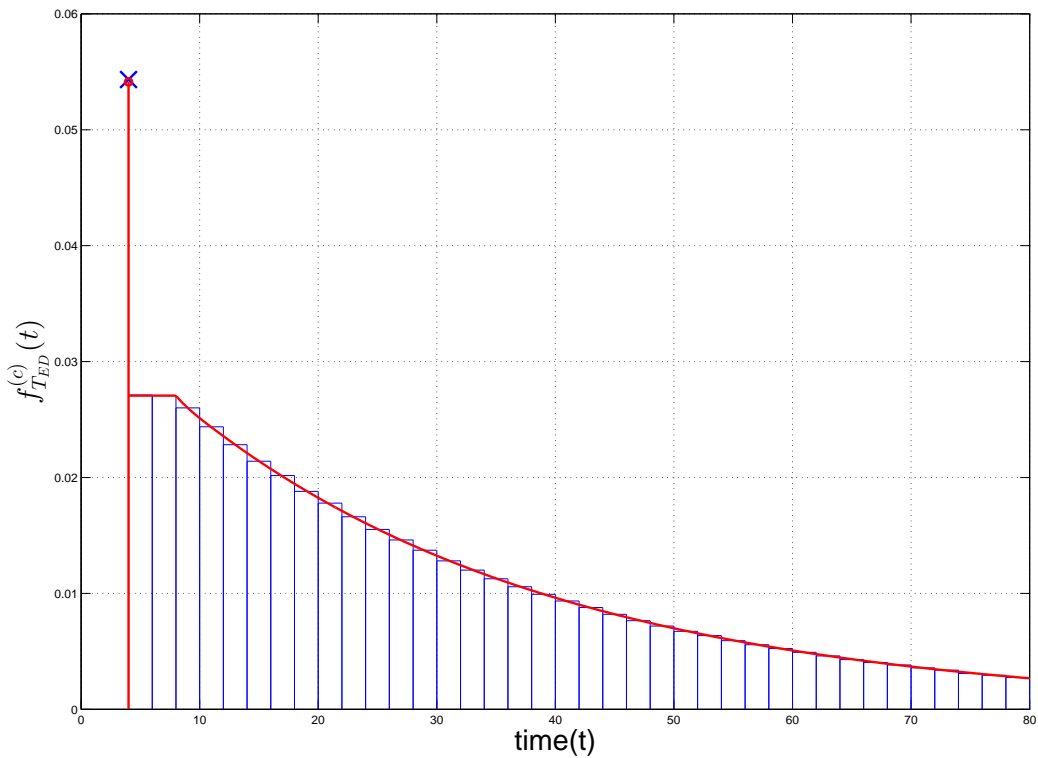}
\caption{Simulation verification for the analytical PDF of $T_{ED}$ with continuous sensing ($T_{tr} = 4$, $\lambda = 3$, and $\mu = 2$).}
\label{lt_cont_sim}
\end{figure}

Fig. \ref{lt_cont_sim} plots the analytical expression for the PDF of the EDT with continuous sensing as given in Eq. (\ref{ft_ed_con}). The corresponding plot for the simulation results is also shown. The perfect match between analytical and simulation results verify our analytical approach.

\subsection{Perfect Periodic Sensing}

For the perfect periodic sensing case, the PDF of the EDT $T_{ED}$ for the SU packet transmission is given by
\begin{equation}
f_{T_{ED}}^{(p)}(t) = \frac{\lambda}{{\lambda}+{\mu}} f_{{T_w},p_{on}}^{(p)}(t-T_{tr}) + \frac{\mu}{{\lambda}+{\mu}} f_{{T_w},p_{off}}^{(p)}(t-T_{tr}),
\label{ft_ed_per}
\end{equation}
where $f_{{T_w},p_{on}}^{(p)}(t)$ and $f_{{T_w},p_{off}}^{(p)}(t)$ denote the PDFs of the waiting time of the SU with perfect periodic sensing, for the case when PU is on at $t=0$, and for the case when PU is off at $t=0$, respectively. We again derive the PDF of waiting time through MGF approach. The MGF of $T_{w,p_{off}}$ for the perfect periodic sensing case, ${\cal{M}}_{{T_w},p_{off}}^{(p)}(s)$, can be calculated as
\begin{align}
{\cal{M}}_{{T_w},p_{off}}^{(p)}(s) = \sum_{k=1}^{\infty} {{\cal{P}}_k} \times \left({\cal{M}}_{T_{wait}}^{(p)}(s) \right)^{k-1} \times \left({\cal{M}}_{T_{waste}}(s) \right)^{k-1},
\label{ftw_poff_per_summation}
\end{align}
where ${{\cal{P}}_k}$ is the probability that the SU was successful in sending the packet in the $k^{th}$ transmission slot, given in Eq. (\ref{P_k}), ${\cal{M}}_{T_{waste}}(s)$ is the MGF of the time duration of a wasted transmission slot $T_{waste}$, which is, noting that the PDF of $T_{waste}$ remains the same as given in Eq. (\ref{ft_mu_tr}) due to the memoryless property of exponential distribution, given in Eq. (\ref{M_ft_mu_tr}), and ${\cal{M}}_{T_{wait}}^{(p)}(s)$ denotes the MGF of the wait time in a single waiting slot. With periodic sensing, $T_{wait}$ consists of multiple $T_s$, and follows a geometric distribution. The MGF can be obtained as
\begin{equation}
{\cal{M}}_{T_{wait}}^{(p)}(s) = \sum_{n=1}^{\infty}(1-\beta) \beta^{n-1} e^{n s T_s},
\label{M_ft_per_lam}
\end{equation}
where $\beta$ denotes the probability that the primary user is on at a given sensing instant provided that it was on at the previous sensing instant $T_s$ time units earlier, given by $\beta=\frac{\lambda}{\lambda+\mu}+{\frac{\mu}{\lambda+\mu}}{e^{-(\frac{1}{\lambda}+\frac{1}{\mu}){T_s}}}$ \cite{Usman}.
Note that $\beta$ is a constant again due to the memoryless property of exponential distribution. Substituting Eqs. (\ref{P_k}), (\ref{M_ft_mu_tr}), and (\ref{M_ft_per_lam}) into Eq. (\ref{ftw_poff_per_summation}), while noting $({\cal{M}}_{T_{wait}}^{(p)}(s))^k = \sum_{n=k}^{\infty}(1-\beta)^k \beta^{n-k} {{n-1} \choose {k-1}} e^{n s T_s}$, we get
\begin{equation}
{\cal{M}}_{{T_w},p_{off}}^{(p)}(s) = e^{-\frac{T_{tr}}{\mu}} + e^{-\frac{T_{tr}}{\mu}} \sum_{k=2}^{\infty} \sum_{n=k-1}^{\infty}(1-\beta)^{k-1} \beta^{n-k+1} {{n-1} \choose {k-2}} e^{n s T_s} \frac{[e^{T_{tr}(s-\frac{1}{\mu})} -1]^{k-1}}{(\mu s - 1)^{k-1}}. 
\end{equation}
After performing some manipulation, using the definition of generalized hypergeometric function, and taking the inverse MGF, we obtain
\begin{align}
\nonumber
& f_{{T_w},p_{off}}^{(p)}(t) = e^{-\frac{T_{tr}}{\mu}} \delta(t) + \sum_{n=1}^{\infty} \left[ \frac{(1-\beta) \beta^{n-1}}{\mu} e^{-\frac{(t - n T_s)}{\mu}} e^{-\frac{T_{tr}}{\mu}} {{}_1 F_1 \left(1-n;1; -\frac{1-\beta}{\beta}\frac{t-n T_s}{\mu} \right)} \right. \\
\nonumber
& \left. + \sum_{i=1}^{n} \left[ (-1)^i e^{-(i+1)\frac{T_{tr}}{\mu}}  {{n-1}\choose{i-1}} \frac{1}{(i-1)!} \frac{(t-i T_{tr} - n T_s)^{i-1}}{\mu^i} (1-\beta)^i \beta^{n-i} e^{\frac{-(t- n T_s - i T_{tr})}{\mu}}  \right. \right. \\
& \left. \left. \times {{}_2 F_2 \left(i+1,i-n;i,i; -\frac{1-\beta}{\beta}\frac{(t-n T_s - i T_{tr})}{\mu} \right)} \right] \right].
\end{align}
Note that the impulse corresponds to the case that the packet is transmitted without waiting.

For the case when PU is on at $t=0$, the MGF of $T_{w,p_{on}}$ for the perfect periodic sensing case, ${\cal{M}}_{{T_w},p_{on}}^{(p)}(s)$ can be calculated as
\begin{align}
{\cal{M}}_{{T_w},p_{on}}^{(p)}(s) = \sum_{k=1}^{\infty} {{\cal{P}}_k} \times \left({\cal{M}}_{T_{wait}}^{(p)}(s) \right)^{k} \times \left({\cal{M}}_{T_{waste}}(s) \right)^{k-1}.
\label{ftw_pon_per_summation}
\end{align}
Substituting Eqs. (\ref{P_k}), (\ref{M_ft_mu_tr}), and (\ref{M_ft_per_lam}) into Eq. (\ref{ftw_pon_per_summation}), and performing similar manipulation as for PU off case, we can arrive at
\begin{equation}
{\cal{M}}_{{T_w},p_{on}}^{(p)}(s) = e^{-\frac{T_{tr}}{\mu}} \sum_{n=1}^{\infty} e^{n s T_s} \beta^n \sum_{i=0}^{n-1} (-1)^i {e^{i T_{tr}(s-\frac{1}{\mu})}} \sum_{k=i+1}^{n} {{k-1}\choose{i}} \left( \frac{1-\beta}{\beta} \right)^{k} {{n-1} \choose {k-1}}  \frac{(-1)^{k-1}}{(\mu s - 1)^{k-1}},
\end{equation}
which, after performing some manipulation, using the definition of hypergeometric function, and taking the inverse MGF, becomes
\begin{align}
\nonumber
& f_{{T_w},p_{on}}^{(p)}(t) = e^{-\frac{T_{tr}}{\mu}} \sum_{n=1}^{\infty} (1 - \beta) \beta^{n-1} \delta(t-n T_s)\\
\nonumber
& + e^{-\frac{T_{tr}}{\mu}} \sum_{n=2}^{\infty} (n-1)  \frac{(1-\beta)^2\beta^{n-2}}{\mu} e^{-\frac{t - n T_s}{\mu}} {{}_1 F_1 \left(2-n;2; -\frac{1-\beta}{\beta} \cdot \frac{t-n T_s}{\mu} \right)} \\
\nonumber
& + \sum_{n=1}^{\infty} \sum_{i=1}^{n-1} (-1)^i e^{-(i+1)\frac{T_{tr}}{\mu}}  {{n-1}\choose{i}} (1-\beta)^{i+1} \beta^{n-i-1} \frac{t^{i-1} e^{-\frac{t-n T_s - i T_{tr}}{\mu}}}{(i-1)! \mu^{i}} \\
& \times {{}_1 F_1 \left(i+1-n;i; -\frac{1-\beta}{\beta} \cdot \frac{t-n T_s - i T_{tr}}{\mu} \right)}.
\end{align}
Note that the sequence of impulses corresponds to the case that the packet is transmitted in the first transmission attempt on acquiring the channel after a random number of sensing intervals/attempts.

\begin{figure}[htb]
\includegraphics[width=6.4 in] {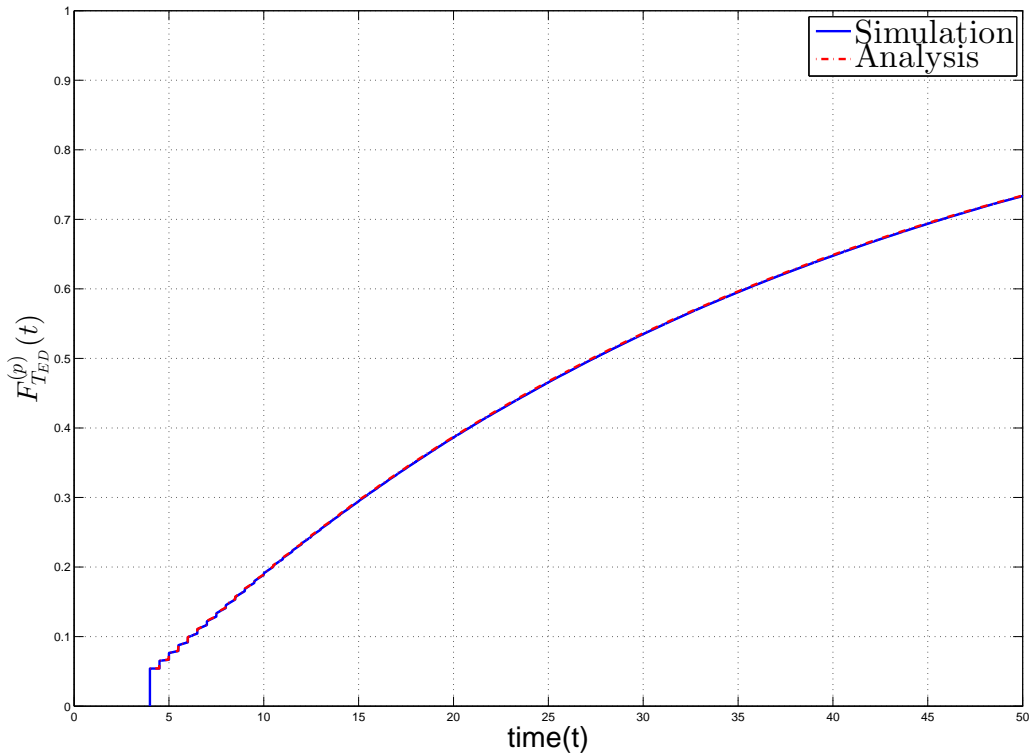}
\caption{Simulation verification for the analytical CDF of $T_{ED}$ with periodic sensing ($T_{tr} = 4$, $\lambda = 3$, $\mu = 2$, and $T_s = 0.5$).}
\label{lt_per_sim}
\end{figure}

Fig. \ref{lt_per_sim} plots the CDF of the EDT with periodic sensing, $F_{T_{ED}}^{(p)}(t)$, obtained by numerical integration of the analytical PDF expression given by Eq. (\ref{ft_ed_per}). The corresponding plot for the simulation results is also shown. The perfect match between analytical and simulation results verify our analytical approach.

\subsection{Imperfect Periodic Sensing}

In the previous section, we assumed that the periodic sensing is perfect, i.e. the SU can always correctly sense whether the channel is free or not. A more practical scenario is with imperfect sensing i.e. the secondary user may not always be able to correctly sense whether the channel is free or not. Specifically, we assume that a busy channel is never sensed as free to protect the PU, while a free channel may be erroneously sensed as busy with a probability $p_e$. We further assume for mathematical tractability that the probability of the primary user turning back on before a successful sensing of idle channel by the secondary user is negligible. Thus, each waiting period of the secondary user can be considered as a sum of two geometric random variables, one catering for the delay until the primary user turns off, and the other accounting for the delay due to missed detection. Denoting the PDFs of the waiting time of the SU with imperfect periodic sensing, for the case when PU is on at $t=0$, and for the case when PU is off at $t=0$, by $f_{{T_w},p_{on}}^{(im)}(t)$ and $f_{{T_w},p_{off}}^{(im)}(t)$, respectively, the PDF of the EDT $T_{ED}$ for the SU is given by
\begin{equation}
f_{T_{ED}}^{(im)}(t) = \frac{\lambda}{{\lambda}+{\mu}} f_{{T_w},p_{on}}^{(im)}(t-T_{tr}) + \frac{\mu}{{\lambda}+{\mu}} f_{{T_w},p_{off}}^{(im)}(t-T_{tr}).
\label{ft_ed_imperfect}
\end{equation}

For the case when PU is off at $t=0$, the MGF of $T_{w,p_{off}}$ for the imperfect periodic sensing case, ${\cal{M}}_{{T_w},p_{off}}^{(im)}(s)$ can be defined as
\begin{align}
{\cal{M}}_{{T_w},p_{off}}^{(im)}(s) = \sum_{k=1}^{\infty} {{\cal{P}}_k} \times \left({\cal{M}}_{T_{wait}}^{(p)}(s) \right)^{k-1} \times \left({\cal{M}}_{T_{mis}}(s) \right)^k \times \left({\cal{M}}_{T_{waste}}(s) \right)^{k-1},
\label{ftw_poff_imperfect_summation}
\end{align}
where $\left({\cal{M}}_{T_{mis}}(s) \right)^k$ is the MGF of the extra waiting time due to sensing errors in $k$ slots, defined by
\begin{equation}
\left({\cal{M}}_{T_{mis}}(s) \right)^k = \sum_{m=0}^{\infty}(1-p_e)^k p_e^{m} {{m+k-1} \choose {k-1}} e^{m s T_s}.
\label{M_ft_per_error_k}
\end{equation}
After substituting Eqs. (\ref{P_k}), (\ref{M_ft_mu_tr}), (\ref{M_ft_per_lam}), and (\ref{M_ft_per_error_k}) into Eq. (\ref{ftw_poff_imperfect_summation}), we arrive at
\begin{align}
\nonumber
{\cal{M}}_{{T_w},p_{off}}^{(im)}(s) = & e^{-\frac{T_{tr}}{\mu}} \sum_{k=1}^{\infty} \sum_{n=k-1}^{\infty}(1-\beta)^{k-1} \beta^{n-k+1} {{n-1} \choose {k-2}} e^{n s T_s} \\
& \sum_{m=0}^{\infty}(1-p_e)^k p_e^{m} {{m+k-1} \choose {k-1}} e^{m s T_s} \frac{1}{(\mu s - 1)^{k-1}} \sum_{i=0}^{k-1} (-1)^{k-i-1} {{k-1}\choose{i}} \cdot {e^{i T_{tr}(s-\frac{1}{\mu})}},
\end{align}
which, on taking the inverse MGF, becomes
\begin{align}
\nonumber
& f_{{T_w},p_{off}}^{(im)}(t) = \sum_{n=1}^{\infty} \sum_{k=2}^{n+1} \sum_{m=0}^{\infty} \sum_{i=0}^{k-1}  (1-\beta)^{k-1} \beta^{n-k+1} (1-p_e)^k p_e^{m} {{n-1} \choose {k-2}} {{m+k-1} \choose {k-1}} {{k-1}\choose{i}} (-1)^{i} \\
& \times e^{-(i+1)\frac{T_{tr}}{\mu}} \frac{(t-n T_s - m T_s - i T_{tr})^{k-2}}{\Gamma[k-1] \mu^{k-1}} e^{-\frac{(t-n T_s - m T_s - i T_{tr})}{\mu}} + \sum_{m=0}^{\infty} (1-p_e) p_e^{m} e^{-\frac{T_{tr}}{\mu}} \delta[t - m T_s].
\end{align}

Similarly, for the case when PU is on at $t=0$, the MGF of $T_{w,p_{on}}$ for the imperfect periodic sensing case, ${\cal{M}}_{{T_w},p_{on}}^{(im)}(s)$ can be defined as
\begin{align}
{\cal{M}}_{{T_w},p_{on}}^{(im)}(s) = \sum_{k=1}^{\infty} {{\cal{P}}_k} \times \left({\cal{M}}_{T_{wait}}^{(p)}(s) \right)^{k} \times \left({\cal{M}}_{T_{mis}}(s) \right)^k \times \left({\cal{M}}_{T_{waste}}(s) \right)^{k-1},
\label{ftw_pon_imperfect_summation}
\end{align}
which, after substituting Eqs. (\ref{P_k}), (\ref{M_ft_mu_tr}), (\ref{M_ft_per_lam}), and (\ref{M_ft_per_error_k}), becomes
\begin{align}
\nonumber
& {\cal{M}}_{{T_w},p_{on}}^{(im)}(s) = e^{-\frac{T_{tr}}{\mu}} \sum_{k=1}^{\infty} \sum_{n=k}^{\infty}(1-\beta)^{k} \beta^{n-k} {{n-1} \choose {k-1}} e^{n s T_s} \\
& \sum_{m=0}^{\infty}(1-p_e)^k p_e^{m} {{m+k-1} \choose {k-1}} e^{m s T_s} \frac{1}{(\mu s - 1)^{k-1}} \sum_{i=0}^{k-1} (-1)^{k-i-1} {{k-1}\choose{i}} \cdot {e^{i T_{tr}(s-\frac{1}{\mu})}}.
\end{align}
Finally, taking the inverse MGF, we get
\begin{align}
\nonumber
f_{{T_w},p_{on}}^{(im)}(t) = & \sum_{n=1}^{\infty} \sum_{k=2}^{n} \sum_{m=0}^{\infty} \sum_{i=0}^{k-1}  (1-\beta)^{k} \beta^{n-k} (1-p_e)^k p_e^{m} {{n-1} \choose {k-1}} {{m+k-1} \choose {k-1}} {{k-1}\choose{i}} \\
\nonumber
& \times (-1)^{i} \cdot e^{-(i+1)\frac{T_{tr}}{\mu}} \frac{(t-n T_s - m T_s - i T_{tr})^{k-2}}{\Gamma[k-1] \mu^{k-1}} e^{-\frac{(t-n T_s - m T_s - i T_{tr})}{\mu}} \\
& + \sum_{n=1}^{\infty} \sum_{m=0}^{\infty} (1-\beta) \beta^{n-1} (1-p_e) p_e^{m} e^{-\frac{T_{tr}}{\mu}} \delta[t - n T_s - m T_s].
\end{align}

\begin{figure}[htb]
\includegraphics[width=6.4 in] {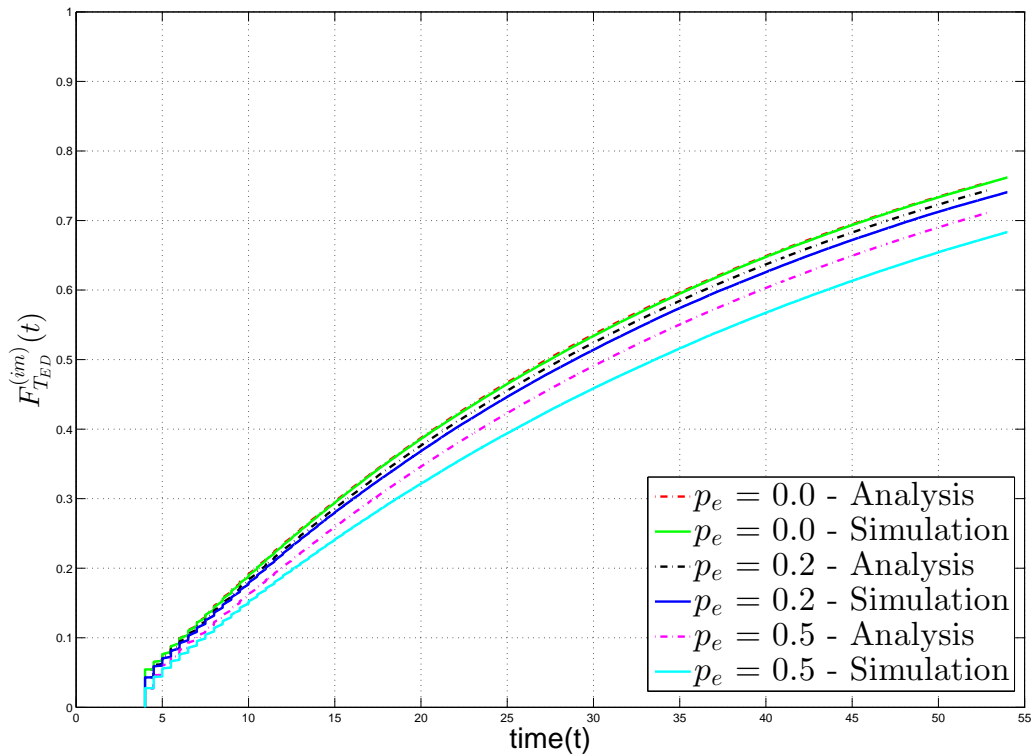}
\caption{Simulation verification for the analytical CDF of $T_{ED}$ with imperfect periodic sensing ($T_{tr} = 4$, $\lambda = 3$, $\mu = 2$, and $T_s = 0.5$).}
\label{lt_imperfect_sim}
\end{figure}

Fig. \ref{lt_imperfect_sim} plots the CDF of the EDT with imperfect periodic sensing, $F_{T_{ED}}^{(im)}(t)$, obtained by numerical integration of the analytical PDF expression given in Eq. (\ref{ft_ed_imperfect}). The corresponding plot for the simulation results is also shown. As can be seen, the analytical results are fairly accurate for small values of $p_e$. The analytical results are slightly different from the simulation results as we ignored the probability that the PU returns before a successful sensing of the idle channel. The plots of the analytical and simulation results for $p_e=0$ match perfectly, which corresponds to the perfect periodic sensing case.

\section{Application to Secondary Queuing Analysis}
\label{SecHT}

In this section, we consider the average transmission delay for the secondary system in a queuing set-up as an application of the analytical results in previous section. In particular, the secondary traffic intensity is high and, as such, a first-in-first-out queue is introduced to hold packets until transmission. We assume that equal-sized packet arrival follows a Poisson process with intensity $\frac{1}{\psi}$, i.e. the average time duration between packet arrivals is ${\psi}$. For the sake of simplicity, the packets are assumed to be of the same length, such that their transmission time $T_{tr}$ be a fixed constant in the following analysis. As such, the secondary packet transmission can be modelled as a general M/G/1 queue, where the service time is closely related to the EDT studied in the previous section. \footnote{The analytical results can also be applied to more complicated queuing models and traffic models, with some further manipulation. They are out of scope of this paper.}

Note also from the EDT analysis, the waiting time of a packet depends on whether the PU is on or off when the packet is available for transmission. As such, different secondary packets will experience two types of service time characteristics. Specifically, some packets might see upon arrival that there are one or more packets waiting in the queue or being transmitted. Such packets will have to wait in the queue until transmission completion of previous packets. Once all the previous packets are transmitted, the new arriving packet will find the PU to be off. We term such packets as type 1 packets. On the other hand, some packets will arrive when the queue is empty, and will immediately become available for transmission. Such packets might find the PU to be on or off. We will call this type of packets, type 2 packets. To facilitate subsequent queuing analysis, we now calculate the first and second moments of the service time for these two types of packets \cite{Usman}. We focus on the perfect periodic sensing case in the following while noting the analysis for the remaining two sensing scenarios can be similarly solved.

\subsection{Service Time Moments}
\subsubsection{First moments}
We first consider the first moment of the service time for packets seeing PU off, denoted by $ST_{p_{off}}$. Noting that $ST_{p_{off}} = T_{w,p_{off}} + T_{tr}$, due to the memoryless property of our scenario with non-work-preserving strategy, we can calculate its mean $E[ST_{p_{off}}]$ by following the conditional expectations appraoch as
\begin{equation}
E[ST_{p_{off}}] = e^{-\frac{T_{tr}}{\mu}} \cdot T_{tr} + (1-e^{-\frac{T_{tr}}{\mu}}) \cdot E[(T_{waste} + T_{wait} + ST_{p_{off}})].
\label{E_off_t_expression}
\end{equation}
Here the first addition term corresponds to the case that the complete packet is successfully transmitted in the first transmission slot, and the second addition term refers to the case when the complete packet is not successfully transmitted. For periodic sensing, it can be shown, from Eqs. (\ref{M_ft_mu_tr}) and (\ref{M_ft_per_lam}), that $E[T_{wait}] = \frac{T_s}{1-\beta}$ and $E[T_{waste}] = \mu - T_{tr} \frac{e^{-\frac{T_{tr}}{\mu}}}{1-e^{-\frac{T_{tr}}{\mu}}}$. The first moment can be calculated from Eq. (\ref{E_off_t_expression}) as
\begin{equation}
E[ST_{p_{off}}] = \frac{1-e^{-\frac{T_{tr}}{\mu}}}{e^{-\frac{T_{tr}}{\mu}}} \left(\mu + \frac{T_s}{1-\beta}\right).
\label{E_off_t}
\end{equation}
Since the case with PU on at $t=0$ is precisely the same as the case of PU off at $t=0$ preceded by a waiting slot, we can define $E[ST_{p_{on}}]$, the first moment of the service time for packets seeing PU on as
\begin{equation}
E[ST_{p_{on}}] = E[ST_{p_{off}}] + \frac{T_s}{1-\beta} = \frac{1-e^{-\frac{T_{tr}}{\mu}}}{e^{-\frac{T_{tr}}{\mu}}} \cdot \mu +  \frac{1}{e^{-\frac{T_{tr}}{\mu}}} \cdot \frac{T_s}{1-\beta}.
\label{E_on_t}
\end{equation}

\subsubsection{Second moments}
Using a similar technique for calculating the first moment, we can write
\begin{equation}
E[ST_{p_{off}}^2] = e^{-\frac{T_{tr}}{\mu}} \cdot T_{tr}^2 + (1-e^{-\frac{T_{tr}}{\mu}}) \cdot E[(T_{waste} + T_{wait} + ST_{p_{off}})^2].
\end{equation}
It can be shown from Eqs. (\ref{M_ft_mu_tr}) and (\ref{M_ft_per_lam}) that $E[T_{wait}^2] = T_s^2 \frac{1+\beta}{(1-\beta)^2}$ and $E[T_{waste}^2] = 2 {\mu}^2 + \frac{e^{-\frac{T_{tr}}{\mu}}}{1-e^{-\frac{T_{tr}}{\mu}}} (-T_{tr}^2 -2 \mu T_{tr})$. Expanding the terms inside $E[.]$, the above equation can be written as
\begin{align}
\nonumber
E[ST_{p_{off}}^2] = & e^{-\frac{T_{tr}}{\mu}} \cdot T_{tr}^2 + (1-e^{-\frac{T_{tr}}{\mu}}) \cdot \left( T_s^2 \frac{1+\beta}{(1-\beta)^2} + 2 {\mu}^2 + \frac{e^{-\frac{T_{tr}}{\mu}}}{1-e^{-\frac{T_{tr}}{\mu}}} (-T_{tr}^2 -2 \mu T_{tr}) + E[ST_{p_{off}}^2]  \right.\\
\nonumber
& \left. + 2 \frac{T_s}{1-\beta} \left(\mu - T_{tr} \frac{e^{-\frac{T_{tr}}{\mu}}}{1-e^{-\frac{T_{tr}}{\mu}}} \right) + 2 \frac{T_s}{1-\beta} \cdot \frac{1-e^{-\frac{T_{tr}}{\mu}}}{e^{-\frac{T_{tr}}{\mu}}} \left(\mu + \frac{T_s}{1-\beta}\right)  \right.\\
& \left. + 2 \left(\mu - T_{tr} \frac{e^{-\frac{T_{tr}}{\mu}}}{1-e^{-\frac{T_{tr}}{\mu}}} \right) \cdot \frac{1-e^{-\frac{T_{tr}}{\mu}}}{e^{-\frac{T_{tr}}{\mu}}} \left(\mu + \frac{T_s}{1-\beta}\right) \right).
\end{align}
Simplifying the above, we obtain
\begin{align}
\nonumber
E[ST_{p_{off}}^2] = & \frac{1}{e^{-\frac{T_{tr}}{\mu}}} \left[-2 T_{tr} \frac{T_s}{1-\beta} - 2 \mu T_{tr} \right] + \left( \frac{1-e^{-\frac{T_{tr}}{\mu}}}{e^{-\frac{T_{tr}}{\mu}}} \right) \frac{1}{e^{-\frac{T_{tr}}{\mu}}} \left[ 2 \mu \frac{T_s}{1-\beta} + 2 \mu^2 \right]  \\
& + \left( \frac{1-e^{-\frac{T_{tr}}{\mu}}}{e^{-\frac{T_{tr}}{\mu}}} \right) T_s^2 \frac{1+\beta}{(1-\beta)^2} + \left( \frac{1-e^{-\frac{T_{tr}}{\mu}}}{e^{-\frac{T_{tr}}{\mu}}} \right)^2 \left[ 2 \frac{T_s^2}{(1-\beta)^2} + 2 \mu \frac{T_s}{1-\beta} \right].
\label{E_off_t2}
\end{align}
Similarly, the second moment of the service time for packets seeing PU on, $E[ST_{p_{on}}^2]$ can be defined as
\begin{equation}
E[ST_{p_{on}}^2] = E[(T_{wait}+ST_{p_{off}})^2],
\end{equation}
which, after substitution of the relevant terms and simplification, becomes
\begin{align}
\nonumber
E[ST_{p_{on}}^2] = & \frac{1}{e^{-\frac{T_{tr}}{\mu}}} \left[-2 T_{tr} \frac{T_s}{1-\beta} - 2 \mu T_{tr} + T_s^2 \frac{1+\beta}{(1-\beta)^2} \right] \\
& + \left( \frac{1-e^{-\frac{T_{tr}}{\mu}}}{e^{-\frac{T_{tr}}{\mu}}} \right) \frac{1}{e^{-\frac{T_{tr}}{\mu}}} \left[ 2 \mu \frac{T_s}{1-\beta} + 2 \mu^2 + 2 \frac{T_s^2}{(1-\beta)^2} + 2 \mu \frac{T_s}{1-\beta} \right]. 
\label{E_on_t2}
\end{align}

\subsection{Queuing Analysis}
The moments of the service time for type 1 packets are same as the moments of those packets which find PU off, i.e.
\begin{equation}
E_1[ST_{type1}] = E[ST_{p_{off}}],
\end{equation}
and
\begin{equation}
E_1[ST_{type1}^2] = E[ST_{p_{off}}^2],
\end{equation}
where $E[ST_{p_{off}}]$ and $E[ST_{p_{off}}^2]$ are defined in Eqs. (\ref{E_off_t}) and (\ref{E_off_t2}), respectively.

As per the analysis given in \cite{Usman}, the moments of the service time for type 2 packets are given by
\begin{equation}
E_1[ST_{type2}] = P_{on,2} \cdot E[ST_{p_{on}}] + (1-P_{on,2}) \cdot E[ST_{p_{off}}],
\label{E2_t}
\end{equation}
and
\begin{equation}
E_1[ST_{type2}^2] = P_{on,2} \cdot E[ST_{p_{on}}^2] + (1-P_{on,2}) \cdot E[ST_{p_{off}}^2],
\label{E2_t2}
\end{equation}
where $P_{on,2}$ denotes the probability that a type 2 packet finds PU on upon arrival, given by
\begin{equation}
P_{on,2} = \frac{\lambda \psi}{\lambda \psi + \lambda \mu + \mu \psi},
\label{P_on_2}
\end{equation}
and $E[ST_{p_{on}}]$ and $E[ST_{p_{on}}^2]$ are defined in Eqs. (\ref{E_on_t}) and (\ref{E_on_t2}), respectively.

Finally,the average total delay for secondary packets can be expressed as \cite{Usman}
\begin{equation}
E[D] = \frac{\psi E_2[t]}{\psi + E_2[t] - E_1[t]} + \frac{ E[t^2]}{2({\psi}-E_1[t])},
\label{E_D}
\end{equation}
and the average number of packets waiting in the queue, not including the packet currently being transmitted, as
\begin{equation}
E[N_Q] = \frac{E[t^2]}{2{\psi}({\psi}-E_1[t])}.
\end{equation}

\begin{figure} 
\includegraphics[width=6.4 in]{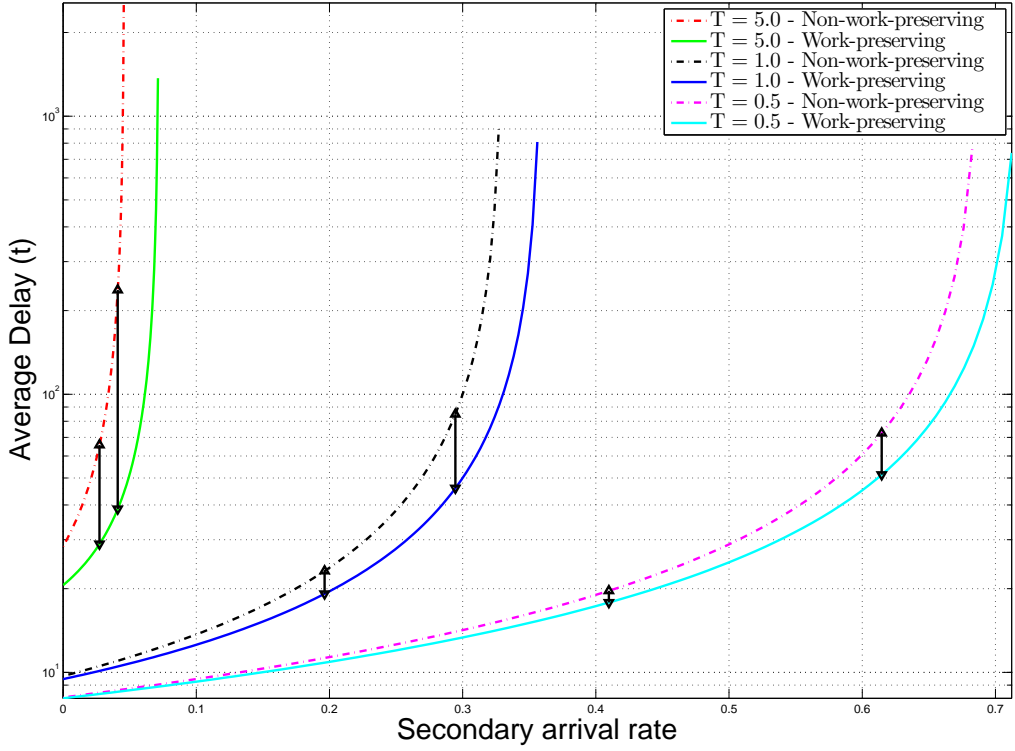}

\caption{Average queuing delay with perfect periodic sensing ($T_s=0.5$, $\lambda = 10$ and $\mu = 6$)}
\label{ht_per_sim}
\end{figure}

Fig. \ref{ht_per_sim} shows the average delay including the queuing delay against the rate of arrival of data packets, for various values of $T_{tr}$, both for work-preserving and non-work-preserving strategies. It can be seen that as expected, work-preserving strategy always performs better than non-work-preserving strategy. Also, the performance difference between the two strategies reduces as the packet transmission time $T_{tr}$ decreases, as shown by the vertical lines in the figure.

\begin{figure} 
\includegraphics[width=6.4 in]{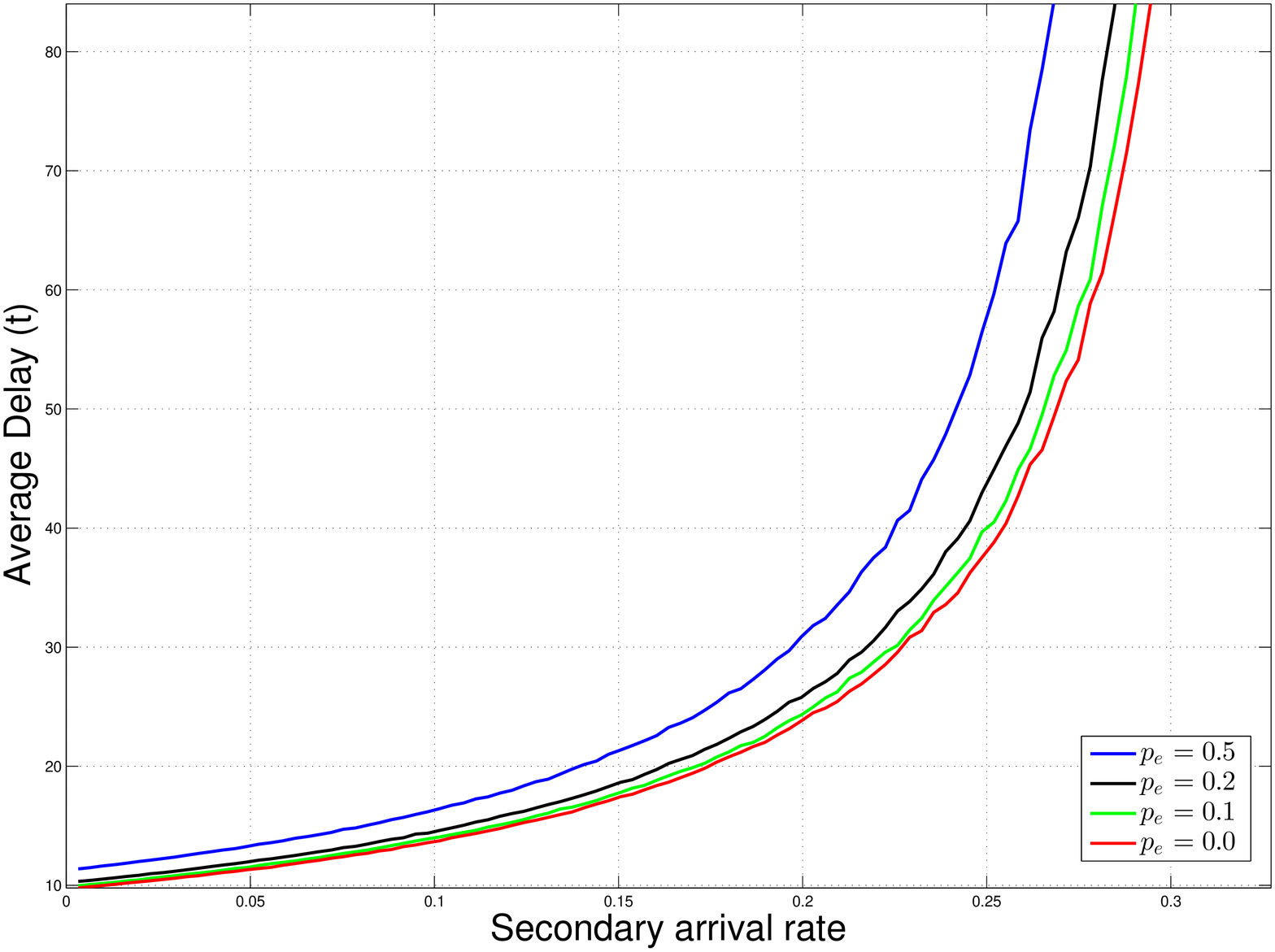}

\caption{Average queuing delay with imperfect periodic sensing ($T_s=0.5$, $\lambda = 10$, $\mu = 6$, and $T=1$)}
\label{ht_imperfect_sim}
\end{figure}

Fig. \ref{ht_imperfect_sim} shows the simulation results for average delay including the queuing delay against the rate of arrival of data packets, for various values of $p_e$, for imperfect periodic sensing scenario. It can be seen that as expected, increasing $p_e$ increases the queuing delay.

\section{Conclusion}
\label{Conclusion}
This paper studied the extended delivery time of a data packet appearing at the secondary user in an interweave cognitive setup assuming non-work-preserving strategy. Exact analytical results for the probability distribution of the EDT for a fixed-size data packet were obtained for continuous sensing, perfect periodic sensing, and imperfect periodic sensing. These results were then applied to analyze the expected delay of a packet at SU in a queuing setup. Simulation results were presented to verify the analytical results. These analytical results will facilitate the design and optimization of secondary systems for diverse target applications. 

\appendix  

In this appendix, we prove by induction, Eq. (\ref{form_partial_fractions}). The base case, with $n=0$, is obvious, as
\begin{equation}
\frac{1}{x(x-a)} =  \frac{1}{a} \left[ - \frac{1}{x} + \frac{1}{x-a} \right].
\end{equation}
Assuming that the equation is true for $n=k$, we can write
\begin{align}
\nonumber
\frac{1}{[x(x-a)]^{k+1}}  &= \frac{1}{[x(x-a)]^k} \times \frac{1}{x(x-a)}\\
& = \sum_{j=0}^{k-1} (-1)^{k} {{2k-j-2} \choose {k-1}} \frac{1}{a^{2k-j-1}} \left[ \frac{1}{x^{j+2} (x-a)} + \frac{(-1)^{j+1}}{(x-a)^{j+2} x} \right].
\label{induction_n_k}
\end{align}
To prove the induction hypothesis, we will use the following two relationships,
\begin{equation}
\frac{1}{x^k(x-a)} = \frac{1}{a^k (x-a)} + \sum_{i=1}^{k} \frac{-1}{a^{k+1-i}x^i},
\label{proposition1}
\end{equation}
and
\begin{equation}
\frac{1}{x(x-a)^k} = \frac{(-1)^k}{a^k x} + \sum_{i=1}^{k} \frac{(-1)^{k-i}}{a^{k+1-i}(x-a)^i}.
\label{proposition2}
\end{equation}
Eq. (\ref{proposition1}) can be proved using the following argument,
\begin{align}
\nonumber
\frac{1}{a^k (x-a)} + \sum_{i=1}^{k} \frac{-1}{a^{k+1-i}x^i} = \frac{1}{a^k (x-a)} \left[ \frac{x^k}{a^k} - \sum_{i=1}^{k} \frac{x^{k-i} (x-a)}{a^{k+1-i}} \right] \\
= \frac{1}{a^k (x-a)} \left[ \frac{x^k}{a^k} - \sum_{i=1}^{k} \frac{x^{k+1-i}}{a^{k+1-i}} + \sum_{i=1}^{k} \frac{x^{k-i}}{a^{k-i}} \right] = \frac{1}{x^k(x-a)}.
\end{align}
Eq. (\ref{proposition2}) can also be proved using a similar argument. Substituting Eqs. (\ref{proposition1}) and (\ref{proposition2}) into Eq. (\ref{induction_n_k}), we obtain
\begin{align}
\nonumber
\frac{1}{[x(x-a)]^{k+1}} = \sum_{j=0}^{k-1} (-1)^{k} {{2k-j-2} \choose {k-1}} \frac{1}{a^{2k-j-1}} & \left[ \sum_{i=1}^{j+2} \frac{-1}{a^{j+3-i} x^i} + \frac{1}{a^{j+2}(x-a)} \right.\\
& \left. + \sum_{i=1}^{j+2} \frac{(-1)^{i+1}}{a^{j+3-i} (x-a)^i} + \frac{-1}{a^{j+2}x} \right].
\end{align}
Performing some further manipulation on Eq. (\ref{induction_n_k}), and using the identity
\begin{equation}
\sum_{k=0}^{n} {{m+k-1} \choose {k}} = {{n+m} \choose {n}},
\end{equation}
it can be shown that
\begin{equation}
\frac{1}{[x(x-a)]^{k+1}} = \sum_{i=0}^{k} (-1)^{k+1} {{2k-i} \choose {k}} \frac{1}{a^{2k-i+1}} \left[ \frac{1}{x^{i+1}} + \frac{(-1)^{i+1}}{(x-a)^{i+1}} \right],
\end{equation}
which proves the induction hypothesis.


\end{document}